\documentclass{emulateapj}
\usepackage{graphicx}
\usepackage[flushleft]{threeparttable}

\slugcomment{In preperation for ApJ Letters}
\shorttitle{[NII] in BR1202-0725}
\shortauthors{Decarli et al.}

\def\Lsun{L$_\odot$}

\def\Hii{H\,{\sc ii}}
\def\Oiii{[O\,{\sc iii}]}

\def\Oi{[O\,{\sc i}]}
\def\Ci{[C\,{\sc i}]}
\def\Nii{[N\,{\sc ii}]}

\def\Cii{[C\,{\sc ii}]}
\def\Ci{[C\,{\sc i}]}

\def\Hii{H\,{\sc ii}}

\def\kms{km\,s$^{-1}$}

\def\lsim{\mathrel{\rlap{\lower 3pt \hbox{$\sim$}} \raise 2.0pt \hbox{$<$}}}
\def\gsim{\mathrel{\rlap{\lower 3pt \hbox{$\sim$}} \raise 2.0pt \hbox{$>$}}}

\begin{document}

\title{
Varying \Cii/\Nii{} line ratios in the interacting system BR1202-0725 at $z=4.7$.
}

\author{
R. Decarli\altaffilmark{1}, 
F. Walter\altaffilmark{1},
C. Carilli\altaffilmark{2},
F. Bertoldi\altaffilmark{3},
P. Cox\altaffilmark{4,5},
C. Ferkinhoff\altaffilmark{6},
B. Groves\altaffilmark{1},
R. Maiolino\altaffilmark{7},
R. Neri\altaffilmark{4},
D. Riechers\altaffilmark{6},
A. Weiss\altaffilmark{8}
}
\altaffiltext{1}{Max-Planck Institut f\"{u}r Astronomie, K\"{o}nigstuhl 17, D-69117, Heidelberg, Germany. E-mail: {\sf decarli@mpia.de}}
\altaffiltext{2}{NRAO, Pete V.\,Domenici Array Science Center, P.O.\, Box O, Socorro, NM, 87801, USA}
\altaffiltext{3}{Argelander Institute for Astronomy, University of Bonn, Auf dem H\"{u}gel 71, 53121 Bonn, Germany}
\altaffiltext{4}{IRAM, 300 rue de la piscine, F-38406 Saint-Martin d'H\`eres, France}
\altaffiltext{5}{ALMA, Chile}
\altaffiltext{6}{Cornell University, 220 Space Sciences Building, Ithaca, NY 14853, USA}
\altaffiltext{7}{Cavendish Laboratory, University of Cambridge, 19 J J Thomson Avenue, Cambridge CB3 0HE, UK}
\altaffiltext{8}{Max-Planck-Institut f\"{u}r Radioastronomie, Auf dem H\"{u}gel 69, 53121 Bonn, Germany}

\begin{abstract}
We study the properties of the interstellar medium in the interacting system BR1202-0725 at $z=4.7$ via its \Nii{} and \Cii{} fine--structure line emission. This system consists of a QSO, a sub-mm galaxy (SMG), and two Ly-$\alpha$ emitters (LAEs). Such a diversity in galaxy properties makes BR1202-0725 a unique laboratory of star formation and galaxy evolution at high redshift. We present ionized nitrogen (\Nii{} 205\,$\mu$m) observations of this system, obtained with the IRAM Plateau de Bure Interferometer. We find no \Nii{} emission at the quasar location, but tentative \Nii{} line detections associated with the SMG and one of the LAEs. Together with available ionized carbon (\Cii{} $158$ $\mu$m) ALMA observations of this system, we find the following: The \Cii/\Nii{} luminosity ratio is $>5.5$ for the QSO and the SMG, but it is as low as $\sim2$ in the LAE, suggesting that, in this source, most of the \Cii{} emission is associated with the ionized medium (\Hii{} regions) rather than the neutral one (PDRs). This study demonstrates the importance of combined studies of multiple fine--structure lines in order to pin down the physical properties of the interstellar medium in distant galaxies. 
\end{abstract}

\section{Introduction}

In local galaxies, optical/UV nebular lines are among the most used diagnostics of the physical properties of the interstellar medium \citep[ISM; see e.g.][]{bpt81,vielleux87,kewley06,groves10}. However, these diagnostics are not accessible at high-$z$, as they are shifted in the mid-infrared regime, where sensitive spectroscopy is out of reach for present instrumentation. The only optical/UV emission line detected at $z>4$ is hydrogen's Ly-$\alpha$, which is hard to interpret in terms of ISM physics, due to its resonant nature \citep[see e.g.][]{yang12}. 

In this context, fine--structure transitions represent key tools to investigate the ISM properties at high-$z$. With rest-frame wavelengths in the 50--500\,$\mu$m range, at $z\gsim4$ these lines are redshifted into the (sub-)mm transparent windows of the atmosphere. The most studied fine--structure line is the ionized carbon (\Cii{}) line at 158\,$\mu$m, which is now (almost) routinely detected in IR--bright sources at high redshift \citep{maiolino05,maiolino09,iono06,walter09a,wagg10,cox11,wagg12,walter12,venemans12,swinbank12,wang13,carilli13a}. Neutral carbon (\Ci{}) has been reported in about 20 sources at high-$z$ \citep[see][]{walter11,alaghband13}. Over the last two years, other fine structure lines have been detected for the first time in high-$z$ galaxies: neutral oxygen \Oi{} \citep[at 63 and 146\,$\mu$m, see][]{coppin12,ferkinhoff13}, doubly-ionized oxygen \Oiii{} \citep[at 52 and 88\,$\mu$m, see][]{ferkinhoff10}, and ionized nitrogen \Nii{} \citep[at 122 and 205\,$\mu$m; see][]{ferkinhoff11,decarli12,combes12,nagao12}. 

Observations of a combination of these lines provide key diagnostics regarding the physical properties of the atomic and ionized phases of the ISM in galaxies \citep[e.g., the hardness of the UV radiation field, gas temperature and density, mass, metallicity;][]{petuchowski93,vanderwerf99}. 
Ionized nitrogen is of particular interest as its ionization potential (14.53 eV) is slightly above that of hydrogen, so \Nii{} emission traces the ionized medium, with the ratio between the two forbidden-line transitions of \Nii{}, at 122 and 205 $\mu$m, being a direct probe of gas density (in the regime between 10 and 3000\,cm$^{-3}$). The 205 $\mu$m \Nii{} transition has critical density and potential of second ionization that are very close to those of \Cii{} ($n_c\approx  45$ cm$^{-3}$ and $E_{\rm 2nd\,ion.}\approx25-30$\,eV, respectively, if electrons are the main collision partners, as in \Hii{} regions). Their flux ratio is thus a diagnostic of the relative abundance of C$^+$ and N$^+$ \citep[e.g.][]{oberst06,walter09b,nagao12,vallini13}, independently of the hardness of the ionization field. Because the ionization potential of carbon is slightly below the one of hydrogen, the \Cii{} emission can arise both from \Hii{} regions and from the neutral outskirts of dense molecular clouds. Therefore, for a given metallicity, the \Cii--to--\Nii{} flux ratio can be used to probe the fraction of carbon emission arising from the ionized medium (where both C$^+$ and N$^+$ are present) vs.~ that arising from the neutral phase \citep[where N$^+$ is suppressed; see, e.g.,][]{oberst06}. 

\begin{figure}
\begin{center}
\includegraphics[width=\columnwidth]{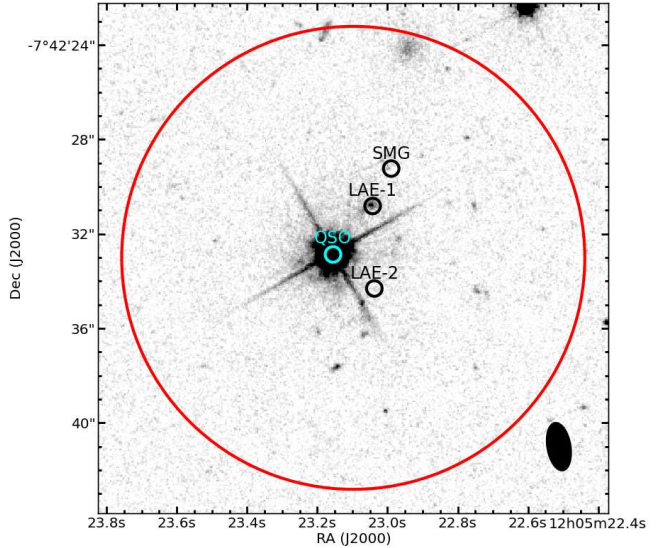}
\end{center}
\caption{Map of BR1202-0725 as observed with {\em HST}/ACS (F775W filter, $i$-band). The various components of this system are labeled. The beam of our \Nii{} observations ($2.0''\times0.9''$) is shown as a black ellipse in the bottom--right corner. The big circle marks the primary beam of PdBI at the observed frequency ($18.4''$).}
\label{fig_ima}
\end{figure}

In this Letter, we present first \Nii{} observations in BR1202-0725, a system of interacting galaxies at $z=4.695$. This system is of particular interest because it encompasses a QSO (South--East), a sub-mm galaxy (SMG, North--West), and two Ly-$\alpha$ emitters (LAEs, one located between the QSO and the SMG, the other located South-West of the QSO), all accommodated within $10''$ on sky ($\approx 65$ kpc at $z=4.695$). Fig.~\ref{fig_ima} shows the {\em Hubble Space Telescope}/Advanced Camera for Survey $i$-band image of the field. Given the diversity of galaxy types in such a small region of the sky, BR1202-0725 represents a unique laboratory of star formation and galaxy assembly only $\sim1.2$ Gyr after the Big Bang.

The QSO and the SMG are among the brightest, unlensed sources in terms of dust luminosity ($L_{\rm IR}>10^{13}$ \Lsun{}). The optical/NIR images of the QSO are completely dominated by the nuclear light, which overwhelms the host galaxy starlight (see Fig.~\ref{fig_ima}). The SMG is barely detected at optical/NIR wavelengths, due to heavy dust obscuration. All the companion galaxies are clearly detected in sensitive {\em HST} and ground--based broad--band observations of the rest--frame UV/optical stellar continuum. The LAEs are also clearly detected in dedicated narrow--band observations encompassing the Ly-$\alpha$ line \citep{hu96} and in spectroscopic observations of this system \citep{ohyama04,williams14}.

Thanks to its enormous IR luminosity, BR1202-0725 has been long studied at (sub-)mm wavelengths. Multiple CO transitions (1-0, 2-1, 4-3, 5-4, 7-6) have been detected \citep{omont96,carilli02,riechers06,salome12}, together with neutral and ionized carbon fine structure lines \citep{iono06,salome12}. \citet{wagg12} and \citet{carilli13a} presented Atacama Large Millimeter Array (ALMA) observations of the \Cii{} emission in BR1202-0725. The \Cii{} emission is clearly detected in the QSO and the SMG. Interestingly, they also reported continuum and tentative line detections associated with LAE--2, although only part of the line width fell into the ALMA bandpass. A \Cii{} detection is also reported at the position of LAE--1. These observational efforts have provided us with a plethora of probes of the neutral and molecular gas in BR1202-0725. The data presented here allow us to complement this information by probing the ionized phase of the ISM through observations of the \Nii{} line.

The Letter is structured as follows: \S\ref{sec_obs} describes our \Nii{} observations of BR1202-0725. \S\ref{sec_results} presents our results, and compares them with the results of \Cii{} studies. Conclusions are drawn in \S\ref{sec_conclusions}. Through the Letter we assume a $\Lambda$CDM cosmology, with $H_0=70$ \kms{}\,Mpc$^{-1}$, $\Omega_{\rm m}=0.3$, and $\Omega_{\Lambda}=0.7$.

\begin{figure}
\begin{center}
\includegraphics[width=\columnwidth]{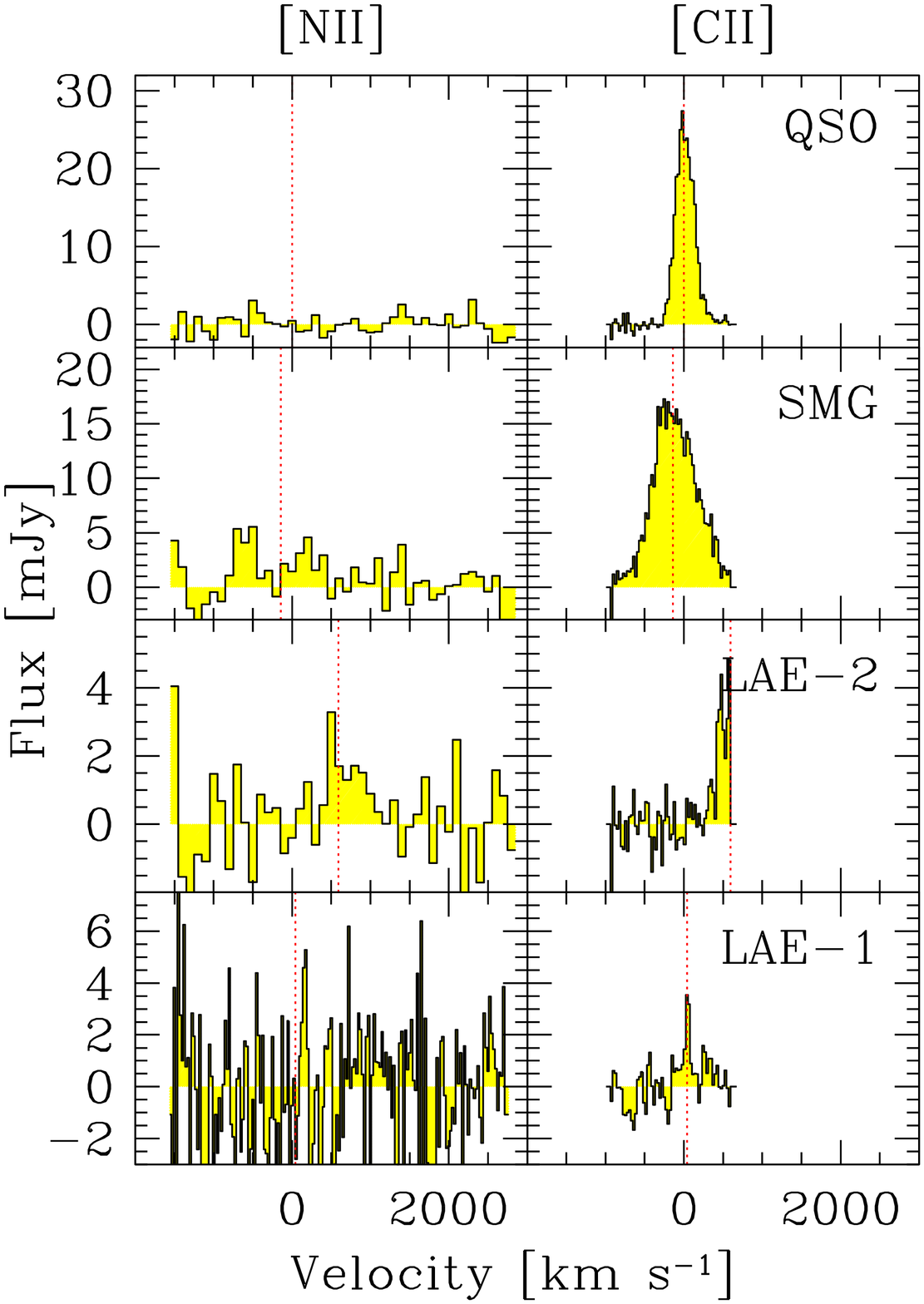}
\end{center}
\caption{Comparison between the continuum-subtracted spectra of \Nii{} (left) and \Cii{} \citep[right;][]{wagg12,carilli13a} in BR1202-0725. These panels have the same scales, to emphasize the wide range of \Cii{}/\Nii{} flux ratios probed with our observations. Vertical lines highlight the peak positions of the lines, as derived by \citet{carilli13a}. No line is detected in the QSO. A tentative, double-horned \Nii{} line detection is found at the position of the SMG. A tentative line is detected at the position of both the Ly-$\alpha$ emitters.}
\label{fig_spc}
\end{figure}

\section{Observations}\label{sec_obs}

We observed BR1202-0725 with the IRAM Plateau de Bure Interferometer (PdBI). We targeted the \Nii{} 205\,$\mu$m transition ($\nu_{\rm 0}=1461.132$ GHz). The tuning frequency was 256.564\,GHz (WideX band 3), corresponding to $z$(\Nii)$= 4.695$. The pointing center was set to RA=12:05:23.100, Dec=-07:42:33.00, J2000.0). The primary beam of PdBI can be described by a Gaussian profile with full width at half maximum (FWHM) = $47.3'' \times (100/\nu)$, where $\nu$ is the observing frequency in GHz. At the tuning frequency of our observations, the primary beam is $18.4''$ in diameter (see Fig.~\ref{fig_ima}). Observations were obtained on January 8, 2013, with the array in compact, 6--antenna configuration (6Cq). Baselines ranged between 18.0 and 176.0\,m. The quasars 3C84 and 1055+018 were observed as flux calibrators. The quasar 3C273 served as phase and amplitude calibrator. We processed our data using the most recent version of the \textsf{GILDAS} software. The receiver operated in the upper sideband, with a typical system temperature of 300\,K. The final cube consists of 5280 visibilities, corresponding to $4.4$\,hr on source (6-antennas equivalent). 

We imaged the cube using the \textsf{GILDAS} suite \textsf{mapping}. Natural weighting was adopted. The beam size is $2.0''\times0.9''$ ($\sim 13$\,kpc $\times6$ kpc at $z=4.695$). We estimate an rms of $1.3$ mJy beam$^{-1}$ per 100 \kms{} channel. We applied Hogbom \textsf{cleaning} down to 2-$\sigma$ per channel in a box encompassing all the individual components of BR1202-0725. 

We compare our \Nii{} observations with the \Cii{} data presented in \citet{wagg12} and \citet{carilli13a}. \Cii{} data were collected with the Atacama Large Millimeter Array (ALMA) during science testing and verification in January 2012. The observations cover 2 GHz of bandwidth, centered at 333.9 GHz, i.e., encompassing the \Cii{} 158\,$\mu$m emission of the QSO and the SMG. The resolution of these data is $1.2''\times 0.8''$. 

\section{Results}\label{sec_results}

\subsection{\Nii{} 205 $\mu$m in BR1202-0725}\label{sec_res1}

Fig.~\ref{fig_spc} shows the continuum--subtracted \Nii{} and \Cii{} spectra of the galaxies in the BR1202-0725 system. The continuum levels have been measured by averaging line--free channels in the cubes (see Table~\ref{tab_lines}), and are consistent with the SED study presented in \citet{wagg14}. No \Nii{} emission is found to be associated with the QSO. A faint \Nii{} line (with possibly two emission peaks) is detected at the position of the SMG. The integrated line flux is $2.7\pm0.5$ Jy\,\kms{}. The line emission is marginally resolved along the West--East axis, consistently with what observed in the CO and \Cii{} emission \citep{salome12,wagg12,carilli13a}. 

LAE--2 shows \Nii{} emission with line flux of $1.10\pm0.35$ Jy\,\kms{}, and a width of 370 \kms{}. The \Nii{} emission from this galaxy matches the line profile observed for the \Cii{} line, although only the blue wing of the \Cii{} line has been covered by the ALMA observations presented in \citet{wagg12} and \citet{carilli13a}. On the other hand, the \Nii{} detection in LAE--1 is only very marginal. The tentative line shows a very narrow profile (FWHM$\approx$50\kms{}) similar to the \Cii{} line, which is however offset in velocity by the same amount. The \Nii{} line flux is $0.28\pm0.12$ Jy\,\kms{}, corresponding to a $\sim2.5$-$\sigma$ nominal detection. Future observations with ALMA are necessary to confirm this tentative detection. Table \ref{tab_lines} summarizes the velocity offsets among the various components and the continuum and line parameters (1461 GHz continuum; width, flux and luminosity of the \Nii{} line) observed in our study.

\begin{table*}
\caption{\rm \Nii{} line parameters in the BR1202-0725 system. (1) Galaxy ID. (2) Continuum flux at 1461 GHz (rest frame). (3) Line peak velocity, with respect to the peak of \Cii{} emission in the QSO (see Fig.~\ref{fig_spc}). (4) \Nii{} 205\,$\mu$m line flux. (5) \Nii{} FWHM. (6--7) \Nii{} line luminosity. (8) \Cii--to--\Nii{} luminosity ratio. (9) IR luminosity, taken from \citet{carilli13a}. Typical uncertainties in the IR luminosities are $\sim20$\%. In case of non-detections, 3-$\sigma$ limits are reported. } \label{tab_lines}
\begin{center}
\begin{tabular}{ccccccccc}
\hline
 Galaxy & $F_{\rm cont}$ & $\Delta v$ & $F_{\rm line}$(\Nii) & FWHM & $L'$(\Nii) & $L$(\Nii) & $L$(\Cii)/$L$(\Nii) & $\log L_{\rm IR}$ \\
        & (mJy)          & (\kms{})   & (Jy\,\kms{})     & (\kms{}) & ($10^9$ K\,\kms{}\,pc$^2$) & ($10^8$ \Lsun{}) &            & (\Lsun{}) \\
    (1) & (2)            & (3)        & (4)    & (5)          & (6)               & (7)       & (8) & (9) \\
\hline
\vspace{0.1cm}
 QSO   & $8.2\pm0.3$ & ---            & $-0.04 \pm 0.31$ &  --- & $ < 4.7$  &  $<4.6     $    &  $>10.3           $ &  13.41  \\
\vspace{0.1cm}
 SMG   & $7.8\pm0.5$ & $-195 \pm 160$ & $ 2.69 \pm 0.53$\footnote{Not corrected for the marginally resolved emission} &  $940_{-300}^{+260}$ & $13.5 \pm 2.6^{\rm a}$ & $ 13\pm3^{\rm a}$ &  $ 5.5_{-1.0}^{+1.5}$ &  13.08  \\
\vspace{0.1cm}
 LAE-1 & $<0.7$      & $ 137 \pm  27$ & $ 0.28 \pm 0.12$ &   $50_{-20}^{+70}$ & $ 1.4 \pm 0.6$ & $1.4\pm0.6$ &  $ 1.0_{-0.5}^{+1.3}$ &  $<11.56$  \\
\vspace{0.1cm}
 LAE-2 & $0.7\pm0.3$ & $ 673 \pm 197$ & $ 1.10 \pm 0.35$ &  $370_{-170}^{+50}$ & $ 5.5 \pm 1.8$ & $5.5\pm1.8$ & --- &  12.23  \\
\vspace{0.1cm}
 LAE-2\footnote{Computed in the velocity range where the \Cii{} is also covered (see Fig.~\ref{fig_spc}).}      & $0.7\pm0.3$ &    ---         & $ 0.39 \pm 0.19$ &  --- & $ 1.9 \pm 0.9$ & $1.9\pm0.9$ & $ 2.3_{-0.9}^{+2.7}$ &  12.23  \\
\hline
\end{tabular}
\end{center}
\end{table*}

\subsection{The \Cii{}/\Nii{} ratio}\label{sec_cii_nii}

\begin{figure}
\begin{center}
\includegraphics[width=\columnwidth]{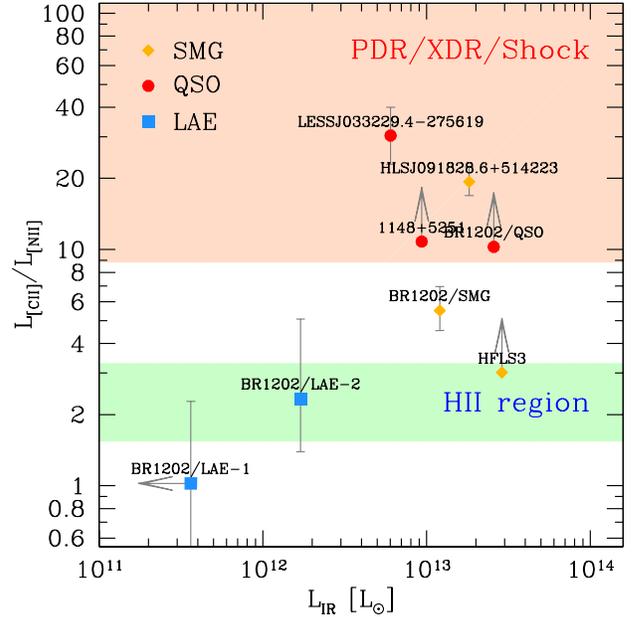}
\end{center}
\caption{The luminosity ratio of \Cii{} and \Nii{} as a function of IR luminosity, as measured in all the high-$z$ galaxies for which \Cii{} has been detected, and sensitive \Nii{} limits are available. Symbols are coded based on the galaxy type. For LAE--2 in BR1202-0725, we have integrated the fluxes of both \Cii{} and \Nii{} only in the velocity range observed in both the lines (see Fig.~\ref{fig_spc}). The ratio for LAE-1 would be a lower limit if follow-up observations were not to confirm our tentative detection.}
\label{fig_hii_pdr}
\end{figure}

\citet{wagg12} and \citet{carilli13a} reported strong \Cii{} detections in the QSO and the SMG. LAE--1 shows a very narrow \Cii{} line (FWHM$\approx$50 \kms{}). \Cii{} line is also reported in LAE--2, though in this case the ALMA observations only covered part of the line profile (see Fig.~\ref{fig_spc}).

The comparison between \Nii{} and \Cii{} emission in the various components of BR1202-0725 reveals highly diverse properties in the properties of the interstellar medium in the various galaxies. We follow \citet{oberst06} in order to estimate the fraction of \Cii{} associated with the ionized medium (stellar \Hii{} regions), in contrast with the neutral surface of dense photon-dominated regions (PDRs) in the outskirt of molecular clouds, or other environments (e.g., X-ray dominated regions, XDRs \citealt{meijerink07}; shocks \citealt{meijerink10,appleton13}; etc). In a \Hii{} region with fixed metallicity, the expected ratio of the two lines is practically independent of the gas density. Assuming $n({\rm N}^+)/n_{\rm e}=7.9\times10^{-5}=0.56 \,n({\rm C}^+)/n_{\rm e}$ \citep{savage96}, we expect a ratio of line luminosities  $L$(\Cii{})/$L$(\Nii{})) $\approx 2$. In the LAEs, the observed \Cii{}--to--\Nii{} ratio is consistent with unity, thus suggesting that these lines are associated with a purely ionized medium. On the other hand, the \Cii{}--to--\Nii{} ratio in the SMG and, most of all, in the QSO, is much higher (see Table \ref{tab_lines}). 

In Fig.~\ref{fig_hii_pdr} we plot the \Cii{}--to--\Nii{} ratio as a function of IR luminosity for the various components of BR1202-0725, and for the few galaxies at $z>1$ for which both \Cii{} has been detected, and deep \Nii{} 205 $\mu$m observations have been reported in the literature \citep{walter09a,combes12,nagao12,riechers13}. These sources are either SMGs or QSO host galaxies with bright IR emission ($L_{\rm IR}\sim 10^{13}$ \Lsun{}). The LAEs in our study have substantially lower IR luminosity ($\sim10^{12}$ \Lsun{} in LAE--2; $<3\times 10^{11}$ \Lsun{} in LAE--1). Moreover, their \Cii{}--to--\Nii{} ratio is substantially lower than typically observed in IR--bright SMGs and QSOs at high--$z$, and is consistent with the range of values expected for \Hii{}--region conditions (shaded green area). On the other side, the majority of the IR--bright sources studied so far have high \Cii{}--to--\Nii{} ratios. This suggests that there is additional \Cii{} emission that may be arising from the {\em neutral} medium in the outskirts of dense molecular regions (PDR--like) or from extended XDRs, where \Nii{} is not present (red shaded area in Fig.~\ref{fig_hii_pdr}). In particular, in the QSO, the observed ratio is $\gsim 3$ times higher than what predicted in the most conservative \Hii{}--region scenario, thus suggesting that $\lsim 33$\% of the \Cii{} in this source is emitted in the ionized phase of the ISM. The global ratio observed in the SMG is lower, $\approx 5.5$, still too high for \Hii{} regions. However, the different profiles of the \Cii{} and \Nii{} lines in this source suggest that the picture is more complex, and that we are probably witnessing a composite source in which both \Hii{} regions and dense molecular clouds play important roles. Deeper data at higher spatial resolution and significantly higher signal-to-noise with ALMA will allow us to perform a velocity- and spatially--resolved study of the \Cii{}--to--\Nii{} ratio, thus disentangling the role of each component.

\section{Conclusions}\label{sec_conclusions}

We have presented first \Nii{} 205\,$\mu$m observations in BR1202-0725, an IR--bright system of interacting galaxies at $z\approx4.7$. The system consists of a QSO, an SMG, and two LAEs. We report \Nii{} detections in the SMG and in the LAE--2. We combine the observations of the \Nii{} with previously published \Cii{} observations. The \Cii{}--to--\Nii{} luminosity ratios span over one order of magnitude, being $<5$ in the LAEs, and $>5$ in the SMG and in the QSO. We use this line ratio to put constraints on the fraction of \Cii{} associated with the {\em ionized} phase of the ISM. The LAEs are well within the parameter space of \Hii{} regions. On the other hand, the \Cii{} emission associated with the neutral phase is high in the SMG and, most of all, in the QSO. This is the first time that the relative importance of the ionized vs neutral phases of the ISM can be directly compared in high-$z$ galaxies.

Deeper observations of this system, in particular using the Atacama Large Millimeter Array, are crucial in order to confirm the tentative \Nii{} detections reported here. In addition, observations of other fine-structure lines (e.g., \Oiii{} and \Oi{}) will allow us to expand our analysis to different regimes (e.g., the highly--ionized medium) and to put first constraints on other parameters (e.g., metallicity) in this unique laboratory of star formation in the early universe.

\section*{Acknowledgments}
This work is based on observations carried out with the IRAM Plateau de Bure Interferometer. IRAM is supported by INSU/CNRS (France), MPG (Germany) and IGN (Spain). This research made use of Astropy, a community-developed core Python package for Astronomy \citep{astropy}. Support for RD was provided by the DFG priority program 1573 ``The physics of the interstellar medium''.

\label{lastpage}

\end{document}